%% file: main.tex
\documentclass[a4paper,10pt]{elsarticle}
\makeatletter
\def\ps@pprintTitle{%
 \let\@oddhead\@empty
 \let\@evenhead\@empty
 \def\@oddfoot{\small{Published in Materials \& Design Volume 259, November 2025, 114889}\hfill}%
 \let\@evenfoot\@oddfoot}
\makeatother

\biboptions{sort&compress} 
\input{parts/packages}

\input{parts/commands}

\title{High heating rate effects in sintering: A phase-field study of La-doped alumina}
\author[1,3]{Marco Seiz}
\ead{marco@kit.ac.jp}
\author[1,2,3]{Tomohiro Takaki}
\ead{takaki@kit.ac.jp}
\affiliation[1]{organization={Faculty of Mechanical Engineering, Kyoto Institute of Technology},
addressline={Matsugasaki, Sakyo-ku},
postcode={606-8585},
city={Kyoto},
country={Japan}}
\affiliation[2]{organization={High-Performance Simulation Research Center, Kyoto Institute of Technology},
addressline={Matsugasaki, Sakyo-ku},
postcode={606-8585},
city={Kyoto},
country={Japan}}
\affiliation[3]{Corresponding authors: marco@kit.ac.jp, takaki@kit.ac.jp}

\begin{document}
\begin{frontmatter}
\begin{abstract}
Sintering is a widespread manufacturing process, accounting for a significant portion of global energy expenditure.
However, controlling this process has been mostly a trial-and-error process, being costly in both time and money.
The recent advance of high heating rate sintering methods, which promise higher energy efficiency and better properties, only adds to this.
This paper aims to reduce these costs by shedding light on the microstructural evolution during high heating rate sintering, which will allow for quicker parameter optimization and improved properties.
The focus will be on how a representative microstructure changes locally as well as globally while resolving grains and the green body at scale, which no prior paper has done.
A representative multiphysics phase-field solver is employed, incorporating a novel particle-based temperature model, which recovered many characteristics typical of high heating rate sintering.
Comparing the simulation data to experimental data showed reasonable agreement over a large density range without parameter adjustment.
Furthermore, the advance of a sintering front including grain growth effects could be shown simulatively for the first time in literature.
These findings suggest that the model can be used for the design of practical heating schedules for the sintering of novel materials.
\end{abstract}

\end{frontmatter}
\section{Introduction}
Sintering is a widespread manufacturing technique which turns a green body, a loosely-packed body of little strength, into a denser, stronger version, shrinking the body in the process.
It is used commercially and industrially for the production of everyday items such as coffee cups, more technical usages such as crucibles for use in casting and even as part of the solar cell manufacturing process\cite{Bucherl2013}.
Key for the latter applications are the materials properties, which depend on the resulting microstructure after the sintering process.
The main microstructural factors for e.g. mechanical strength are the remaining porosity and where it is located as well as the grain size.
High strength is usually obtained by a perfectly dense body with a small grain size.
However, conventional sintering methods expose the body to high temperatures over long times in order to reach a high density.
Over this entire time grain growth can take place, worsening the properties.
On top of this, heating a furnace for hours is also wasteful in terms of energy usage.
With the challenges posed by climate change, improving energy efficiency and potentially also materials properties is an attractive proposition.

In recent years, so-called high heating rate sintering methods\cite{Wang2020,Porz2022a,Gomez2016} have been explored to achieve these goals.
As the name implies, the body is heated up rapidly, likely reducing the energy usage.
If densification has a higher activation energy than grain growth, then moving quickly over the ``low'' temperature regime in which primarily grain growth is active will also yield smaller grains at equivalent density.
However, due to the high heat input, process control becomes even more important than in conventional sintering:
If too much heat is applied overall, grain growth will happen just the same; if too much heat is applied at once, thermal stresses can crack the body.
Furthermore, many heating methods heat only the surface of the body, with thermal diffusion then slowly heating the rest.
This does not only imply thermal stresses, but also potentially differential densification, which can lead to even higher stresses as well as de-sintering\cite{Lange1996b}.

While experimental trial and error with sophisticated monitoring can yield usable process parameters, this can become expensive in both monetary and temporal cost.
An alternative way to optimize the process is to conduct simulations, given that a model approximating the real process exists.
In this paper, such a model is proposed for high heating rate sintering based on the ``fast firing'' process, in which the green body is quickly inserted into a furnace at the target temperature.
This model will be probed whether it is suitable to approximate the process as well as to attack a few basic questions about the high heating rate sintering process:
\begin{itemize}
 \item When does the temperature inhomogeneity become significant?
 \item When does temperature inhomogeneity lead to microstructural inhomogeneity?
 \item Does microstructural inhomogeneity persist once it occurs?
\end{itemize}

Previous modeling and simulation work on high heating rate sintering while spatially resolving both the temperature field and the microstructure is limited to \cite{Teixeira2021,Yang2019b}.
Teixeira et al. \cite{Teixeira2021} constructed a discrete element method (DEM), combining classical DEM with sintering laws and thermal evolution by considering convection, radiation and conduction.
They could show good agreement for the temperature evolution when compared to experimental data.
In contrast to a phase-field approach, they could resolve larger volumes and spans of time, but by virtue of method both grain growth and grain deformation are difficult to include (but see \cite{Paredes-Goyes2021} for a DEM approach for including grain growth).
Yang et al. \cite{Yang2019b} constructed a non-isothermal phase-field model of selective laser sintering including melting of the grains.
The thermal model included conduction, latent heat as well as the laser heat source.
They could show the strong localization of heating effects during selective laser sintering and qualitative agreement with experimental observations.
The energy functional employed has the problems described by \cite{Yang2023} and the lack of advective terms for the microstructural evolution imply a size effect\cite{Seiz2023b,Seiz2023a}, so data relating to densification should be regarded with care\cor{, as it will have a significant effect from the simulation domain size}.
While there are many works simulating the temperature distribution during various kinds of sintering methods on a part scale without resolving the microstructure, resolving both the temperature and microstructure spatially is as-of-yet not well explored territory, which this paper aims to shed light on.

\section{Model and implementation}
\input{parts/model}

\subsection{Parameters}
The parameters are generally based on literature data for alumina. Initial tests with grain mobility of pure alumina showed comparatively rapid grain growth, sometimes leaving only a few grains at higher densities which raises questions of statistical accuracy.
Thus for the grain mobility the La-doped grain mobility found by \cite{Behera2016} is employed.
\cor{This mobility is about a tenth of that of pure alumina, with the reason most likely being solute drag.
Directly incorporating solute drag in the model and using the parameters of alumina is possible, see e.g. the phase-field work on solute drag of \cite{Abdeljawad2017,Guin2024}.
For simplicity and lack of experimental data of La diffusion in alumina, the present paper does not explicitly model the solute drag effect of La, but only its effect of a reduced grain mobility.
}
Strictly speaking, the diffusivities should be affected by the La-doping as well; however, due to the lack of temperature dependent data in the literature, the values for pure alumina were employed.

The temperature $T$ dependence of the diffusion coefficients are modeled with Arrhenius functions $D(T) = D_0 \exp(-\frac{E_a}{RT})$ with the parameters $D_0$ and $E_a$, as well as the ideal gas constant $R$.
For simplicity, the bulk $D_b$ and vapour $D_v$ diffusion coefficients are taken to be equal.
Since the initial compositions of the grains and vapour are in approximate equilibrium, there should be negligible mass transfer between the phases and hence $D_v$ only plays a minor role.
For the conversion of $D_{gb}\delta_{gb}$ to the usual diffusivity units a grain boundary width of $\delta_{gb} = \SI{0.6}{nm}$\cite{Dillon2007} is assumed.
For the relaxation time $t_r$, a convergence study using a two particle geometry was conducted at two temperatures.
A value for $t_r$ was arbitrarily picked and afterwards decreased until the velocity-time curves showed no more significant differences, i.e. achieved convergence.
An Arrhenius function was then calculated based on the results.
This ensures that the densification is controlled by diffusion and not the sink efficiency of the grain boundary.
The surface mobility $M_{s}$ is calculated based on the largest diffusion coefficient as to ensure that the surface relaxation is controlled by diffusion and not the phase-field.

The resulting functions are tabulated in \cref{tab:Tdep} using 3 significant digits; the full values are available in the supplementary material.
The thermodynamic coefficients for \cref{eq:free-energy} are as follows:
The equilibrium vapour concentration $c_v^{eq}$ is 0.02, the equilibrium solid concentration $c_s^{eq}$ is 0.98 and the shared prefactor $k_v = k_s$ is \SI{5e9}{J/m^3}.
The prefactor is chosen high enough such that spontaneous shrinkage due to capillarity \cite{Yue2007} is insignificant.

The timestep is calculated based on a maximum eigenvalue estimate of the respective equations \cref{eq:pfeq,eq:ceq} with a safety factor of \num{0.2}.
If this timestep is larger than the time required for the average temperature to change by more than \SI{0.25}{K}, then the latter time is used for the timestep.
This is done in order to account for the vanishingly small mobilities at the simulation start as well as to keep the initial error on the temperature evolution small.

\begin{table}[h]
\centering
\caption{Parameters and their sources}
\label{tab:params}
 \begin{tabular}{lll}
 parameter   & function & source \\
\hline\\
surface energy $\gamma_s$ & \SI{2.56}{J/m^2} - \SI{0.783e-3}{J/(m^2K) T} & \cite{Nikolopoulos1985a} \\
GB energy $\gamma_{gb}$ & \SI{1.91}{J/m^2} - \SI{0.611e-3}{J/(m^2K) T} & \cite{Nikolopoulos1985a} \\
bulk diffusion $D_b$ & \SI{0.32}{m^2/s} $\exp(-\frac{\SI{653e3}{J/mol}}{RT}) $ & \cite{Heuer2008} \\
vapor diffusion $D_v$ & $D_b$ & - \\
surface diffusion $D_s$ & \SI{7.2}{m^2/s} $\exp(-\frac{\SI{314e3}{J/mol}}{RT}) $  & \cite{Robertson1966} \\
GB diffusion $D_{gb}$ & \SI{9.17e10}{m^2/s} $\exp(-\frac{\SI{825e3}{J/mol}}{RT}) $ & \cite{Heuer2008} \\
GB mobility $M_{gb}$ & \SI{2e-3}{m^4/(Js)} $\exp(-\frac{\SI{443e3}{J/mol}}{RT}) $ & \cite{Behera2016} \\
surface mobility $M_s$ & $\frac{1.1}{\gamma_s} \mathrm{max}(D(T))$ & - \\
relaxation time $t_r$ & \SI{1e-20}{s} $\exp(\frac{\SI{428e3}{J/mol}}{RT}) $ & - \\
heat capacity $c_p$ & $[\num{1.11e3} + \num{0.14}\frac{T}{K} - 411\exp(-0.006\frac{T}{K})]\si{{J/kg}}$ & \cite{Munro1997}\\
heat conductivity $\kappa$ &[$\num{5.85} + \frac{\SI{15.3e3}{K}}{T+\SI{516}{K}} \exp(-0.002\frac{T}{K}) $]\si{J/(Kms)} & \cite{Munro1997}\\
convective coefficient $h$ & \SI{5}{J/(Km^2s)} & \cite{Teixeira2021} \\
emissivity $\epsilon$ & 0.8 & \cite{Teixeira2021}\\
grid spacing $\Delta x$ & \SI{1}{nm} & - \\
interface width $W$ & $6\Delta x$ & - \\
\end{tabular}
\label{tab:Tdep}
\end{table}

\section{Results}
\input{parts/results}

\section{Conclusion}
The influence of spatially homogeneous as well as inhomogeneous heating of a green body on its microstructural evolution was explored.
For this, a previous phase-field model of sintering was extended to include the temperature dependence of its parameters.
Furthermore, a novel and efficient coarse-grained temperature model was developed and implemented to account for the natural multiscale problem presented by sintering.
The model bounded a large regime of the density-densification trajectory of one experimental result, suggesting the applicability of the model for real systems.

The questions posed in the introduction can be answered as follows:
The conditions when the temperature inhomogeneity in a porous, geometrically complex and evolving green body becomes significant can be predicted well with the Biot number $\Bi{} > 0.017$.
Once temperature inhomogeneity becomes significant, microstructural inhomogeneity quickly follows.
However, the inhomogeneity need not necessarily persist.
If the heat transport inside the body is sufficiently fast, densification on the inside can be activated and match the evolution of the microstructure further outside, i.e. a sintering front emerges.
From this one can conclude that for sufficiently small bodies, final homogeneous microstructures can be achieved, even if they temporarily exhibit significant temperature and microstructural inhomogeneity.
The conditions for the successful formation of a sintering front depends on the size of the green body, the heating technique, thermal properties as well as factors affecting the densification and grain growth processes.
The determination of these conditions would help high heating rate sintering technique development significantly, as it would significantly reduce the effort in determining parameters for the often desired homogeneous microstructure.

Based on these results, it is concluded that the developed model is capable of approximating high-heating rate sintering and predicting the microstructural evolution of experiments, subject to materials parameter adjustments.
Future work will include investigation into more efficient time integrators as to access lower heating rates and larger grain sizes with acceptable computational effort.
Furthermore, the effect of the compressive velocity profile across grain boundaries will be investigated further and possible avenues for its elimination will be explored, such that the model can also be applied to lower heating rates.

\section*{Author contributions}
\textbf{Marco Seiz}: Conceptualization, Software, Methodology, Investigation, Data Curation, Validation, Visualization, Writing - original draft, Writing - review \& editing.
\textbf{Tomohiro Takaki}: Funding acquisition, Project administration, Resources, Writing - review \& editing.

\section*{Conflicts of interest or competing interests}
The authors declare that there are no conflicts of interest.

\section*{Data and code availability}
The code required to reproduce the present work cannot be shared publicly.
The processed data is available within the supplementary material.

\section*{Supplementary Material}
The supplementary material of this paper is available at \url{https://doi.org/10.5281/zenodo.15605646}.
It contains interactive and static versions of notebooks reproducing the graphs in the paper, estimates of the Biot number in sintering, the parameters in the employed precision and videos of the simulations.

\section*{Acknowledgements}
\input{parts/acks}

\bibliography{literature}

\end{document}

%% file: parts/packages.tex
\usepackage[english]{babel}
\usepackage[utf8]{inputenc}
\usepackage{graphicx}
\usepackage{caption}
\usepackage[labelformat=simple]{subcaption}

\usepackage{mathtools}
\usepackage{amsfonts}
\usepackage{xfrac}
\usepackage{siunitx}
\usepackage{placeins}
\usepackage{tikz}
\usepackage{pgfplots}
\usetikzlibrary{calc}
\usetikzlibrary{backgrounds}
\usepgflibrary{shapes}
\usetikzlibrary{through}
\usetikzlibrary{intersections}
\usetikzlibrary{matrix}
\usepackage{multirow}
\usepackage{nameref}
\usetikzlibrary {arrows.meta}

\usepackage{hyperref}
\usepackage{cleveref}
\crefname{equation}{eq.}{eqs.}
\crefname{figure}{fig.}{figs.}
\Crefname{figure}{Figure}{Figures}

\crefname{table}{Table}{Tables}
\Crefname{table}{Table}{Tables}

\usepackage{xifthen}
\usepackage{wrapfig}

\usepackage{esdiff}
\usepackage{ifthen}

\usepackage{booktabs}
\usepackage{nicefrac}

\usepackage[
]{todonotes}
\presetkeys{todonotes}{inline}{}
\usepackage{footmisc}

\usepackage[singlespacing]{setspace}

%% file: parts/commands.tex
\definecolor{shadecolor}{RGB}{255,255,0}
\newcommand{\cor}[1]{#1}
\newcommand{\Bi}{\ensuremath{\mathrm{Bi}}}

\newcommand{\pdiff}[2]{\ensuremath{\frac{\partial #1}{\partial #2}}}

\newcommand{\ddiff}[2]{\frac{d #1}{d#2}}     

\newcommand{\grad}[1]{\ensuremath{\nabla{#1}}}

\newcommand{\vv}[1]{\boldsymbol{#1}}

\newcommand{\contact}{\ensuremath{\mathrm{C}}}

\newcommand{\x}{\ensuremath{\vv{\\newcommand{\contact}{\ensuremath{\mathrm{C}}}xi}}}

\newcommand{\vardiff}[2]{\frac{\delta{#1}}{\delta {#2}}}     

\newdimen\CdotAxis
\newcommand*{\CdotAux}[3]{%
  {%
    \settoheight\CdotAxis{$#2\vcenter{}$}%
    \sbox0{%
      \raisebox\CdotAxis{%
        \scalebox{#1}{%
          \raisebox{-\CdotAxis}{%
            $\mathsurround=0pt #2#3$%
          }%
        }%
      }%
    }%
    \dp0=0pt %
    \sbox2{$#2\bullet$}%
    \ifdim\ht2<\ht0 %
      \ht0=\ht2 %
    \fi
    \sbox2{$\mathsurround=0pt #2#3$}%
    \hbox to \wd2{\hss\usebox{0}\hss}%
  }%
}

\def\addlegendimage{\csname pgfplots@addlegendimage\endcsname}

\DeclareSIUnit{\molpc}{mol\text{-}\%}



%% file: parts/model.tex
The model description is split into three parts:
First, the overall phase-field model for representing and evolving the microstructure is described.
Second, the model used to calculate the grain motion is described.
Finally, the models used for the temperature evolution are described.
Afterwards, some notes on computational and measurement aspects are given and the parameter choices are described in the final subsection.

\cor{Note that the model applies to any material that sinters by vacancy absorption, and thus the large material groups of metals and ceramics are covered by the model.
All the following models are agnostic w.r.t. the number of components in the system.
The only necessary condition for the motion model is that some identification of the vacancy concentration is possible such that the vacancy annihilation rate can be estimated.
Furthermore, different solid phases may also be considered to e.g. model co-sintering of layered structures.
This will generally require additional components to account for the chemistry of the different solids; the model itself remains unchanged.
}

\subsection{Phase-field model}
In this work a multi-phase-field (MPF) model based is coupled with a Kim-Kim-Suzuki (KKS) model for the concentration evolution.
The geometry of the microstructure is described with the help of phase-fields, which allow for a geometrically unrestricted description of free-boundary problems.
At each point in space, a phase-field vector $\vec{\phi}$ describes the local volume fractions of different phases.
For solid-state sintering of a pure material, a vapour phase $\phi_V$ and arbitrary many grain phases $\phi_\alpha$ are distinguished.
Since the number of grains needs to be sufficiently large, storing the entire phase-field vector per cell would require too much memory and computational power.
Instead, only the set of non-zero phase-fields along with their phase-indices $\vec{\phi^i}$ is stored, up to a maximum of 8 entries per cell.
This is permissible as the non-trivial phase-field values are concentrated in thin regions called the interface.
This approach is based on \cite{Kim2006,Takaki2009}.

The $n$ phase-fields are evolved following the approach of \cite{Steinbach1999}, in which the sum of pairwise variational derivatives gives the temporal change, combined with a rigid-body advection term:
\begin{align}
 \pdiff{\phi_i}{t} &= - \sum_{j=0}^{n-1} \frac{2L_{ij}(T)}{n} (\vardiff{F}{\phi_i} - \vardiff{F}{\phi_j}) - \nabla \cdot (v_i\phi_i)
 \label{eq:pfeq}
\end{align}
with the energy functional
\begin{align}
 F &= \int_V \sum_{i=0}^{n-1} \sum_{j=i+1}^{n-1} \left[ a(\phi, \nabla \phi, T)+w(\phi, T) \right] + \phi_i f_i(c_i) dV \\
 a(\phi, \nabla \phi, T) &= -\frac{A_{ij}^2}{2}(T) \nabla \phi_i \nabla \phi_j\\
 w(\phi, \nabla \phi, T) &= W_{ij}(T) \phi_i \phi_j\\
 f_i(c_i) &= \frac{k_i}{2} (c_i - c_i^{eq})^2 \label{eq:free-energy}
\end{align}
describing the energy of the entire domain.
The terms $a$ and $w$ describe the equilibrium profile and account for capillary forces, whereas $f$ describes thermodynamic driving forces.
The coefficients $L_{ij}$, $A_{ij}$ and $W_{ij}$ can be related to materials parameters (e.g. surface energy, grain mobility) as described in \cite{Takaki2009}.

In order to decouple the driving force description from the equilibrium conditions, such that the cited relations hold, the concentration evolution is evolved following \cite{Kim1999} with an additional advective term:
\begin{align}
 \pdiff{c}{t} &= \nabla \cdot (D(\phi, T) (\phi_s \nabla c_s + \phi_V \nabla c_V) + c\vec{v}(x)) \label{eq:ceq}\\
 D(\phi, T) &= \sum_{i=0}^{n-1} D_i\phi_i(T) + g(\phi_v, \phi_s) D_s(T) + \sum_{\alpha=1}^{n-1} \sum_{\beta=\alpha+1}^{n-1} g(\phi_\alpha, \phi_\beta) D_{gb}(T)
\end{align}
in which the subscript $s$ refers to the effective solid phase, i.e. $\phi_s = \sum_{\alpha \neq V} \phi_\alpha$.
The diffusion coefficient has contributions from bulk diffusion $D_i$, surface diffusion $D_s$ and grain boundaries $D_{gb}$, with the indices $\alpha, \beta$ only going over the grain phases.

\subsection{Motion model}
The model for the calculation of the grain velocities $v_\alpha$ is based on \cite{Seiz2023b,Seiz2024}.
The idea is to determine the vacancy absorption rate at a grain boundary, which is geometrically connected to the velocity jump across the grain boundary.
The vacancy absorption rate is formulated as
\begin{align}
 \ddiff{n}{t} = \frac{n-n_{eq}(\mu_\alpha\beta)}{t_r}
 \label{eq:vacabsorb}
\end{align}
i.e. a simple relaxation to an equilibrium value $n_{eq}(\mu_\alpha\beta)$ which depends on the average capillary pressure $\mu_{\alpha\beta}$ experienced by both particles.
The conversion from number density $n$ to molar fraction $c$ (concentration) is computed as $n=\frac{N_A}{V_m}c$ with Avogadro's constant $N_A$ and the molar volume $V_m$.
The average capillary pressure is approximated by the average of the chemical potential on a particle's surface.
Following \cite{Seiz2023b}, the velocity jump $dv$ across a grain boundary can be written as
\begin{align}
 dv &= \ddiff{n}{t} \frac{\Omega}{A}
\end{align}
with the atomic volume $\Omega$ and the grain boundary area $A$.
Assuming that all velocity jumps $\Delta v_{ij} = v_i - v_j$ should be fulfilled simultaneously, with justification given in \cite{Seiz2023b}, one arrives at a system of equations
\begin{align}
 \contact{} v = \Delta v
 \label{eq:velsystem}
\end{align}
with the contact matrix $\contact{}$ describing the connectivity of the grains.
This system is typically overdetermined and hence a least-squares solution satisfying conservation of momentum is sought.

Once the particle velocities $v$ have been determined, the velocity field $v(x)$ for the advection of the concentration and vapour phase-field is interpolated in a rigid-body fashion as
\begin{align}
 \vec{v}(x) = \frac{\sum_{\alpha=1}^{n-1} \phi_\alpha \vec{v}_\alpha }{\sum_{\alpha=1}^{n-1} \phi_\alpha }
\end{align}
with $\alpha$ only going over non-vapour phase-fields.
This yields a velocity field of uniform level within grains, smooth transitions across grain boundaries and abrupt jumps to zero velocity in the vapour.

\cor{Note that the effect of particle size is naturally accounted for in the model, as it modifies the local capillary pressure, which in turn changes the chemical potential and thus the fluxes of \cref{eq:ceq}.
These fluxes limit the vacancy absorption rate \cref{eq:vacabsorb}, with the absorption rate also being dependent on the capillary pressure on a particle scale.
The phase-field equation \cref{eq:pfeq} is known to reproduce ideal curvature-driven grain growth and thus also naturally incorporates particle size effects.
For an explicit investigation of the influence of particle size under isothermal conditions, the interested reader is referred to \cite{Seiz2023a,Seiz2023b}.
}

\subsection{Temperature model}
\begin{figure}[h]
\includegraphics[width=\textwidth]{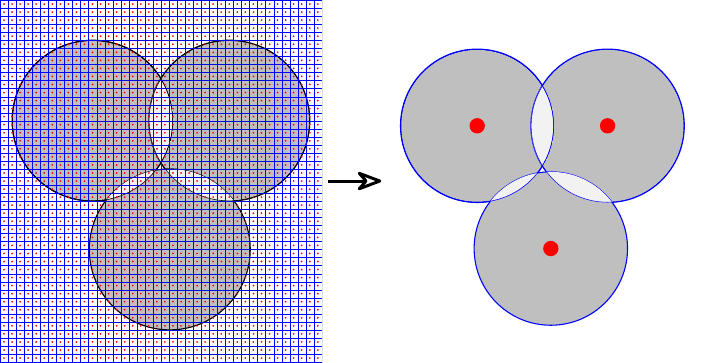}
 \caption{Transformation from grid-wise temperature to particle-wise temperature. Red dots indicate a degree of freedom and blue lines indicate the degree's boundaries.}
 \label{fig:gridp}
\end{figure}

With sintering being a thermally activated process, it is of great importance to resolve the spatiotemporal evolution of the temperature.
However, since the thermal diffusivity is typically several orders of magnitude larger than the mass diffusivity, resolving both thermal and mass diffusion at the same time can become computationally expensive.
Furthermore, for the present work the heating contributions of both convective and radiative heating shall be included, as to model the ``dunk cycle''\cite{German2014} or ``fast firing''\cite{Garcia2011} variant of high heating rate sintering.
This would imply solving a fluid-flow problem for the convective part and a radiative heat transfer problem, both of which are highly computationally expensive.
In order to avoid this expense, a simplified thermal model on the particle scale is proposed, similar to the discrete element method of \cite{Teixeira2021}.
The starting point is the energy balance
\begin{align}
 c_p\pdiff{T}{t} &= -\nabla \cdot (-\kappa \grad T) + q\\
 \int_V c_p \pdiff{T}{t} dV &= -\int_A -\kappa \grad T d\vec{A} + \int_V q dV
\end{align}
in which heat diffusion and a heat source $q$ appear, with the second line converting into the weak form and applying the divergence theorem.
Suppose now that we discretize the space occupied by all grains into individual volumes $V_i$ according to the extent of the grains, then we obtain for each grain
\begin{align}
\int_{V_i} c_{p,i} \pdiff{T_i}{t} dV &= -\sum_{j} \int_{A_{ij}} -\kappa_{ij} (\grad T)_{ij} d\vec{A} + \int_{V_i} q_i dV
\end{align}
which changes the number of unknowns from the grid used to resolve \cref{eq:pfeq,eq:ceq} to the number of particles in the domain, which is typically far smaller.
The index $j$ here goes over all particles to which $i$ is connected, similar to the contact matrix $\contact$ from the motion model.
This procedure is visualized in \cref{fig:gridp} for a three-particle configuration.
There is an implicit no-flux boundary condition on the particle boundaries (blue), as long as these do not intersect with other particles (light gray regions).
The light gray regions describe the contact areas $A_{ij}$ which scales the flux between the particles and are equivalent to the GB areas already being determined in the motion model.

If we assume further that the particle temperature is homogeneous, we obtain after division by $V_i$
\begin{align}
 c_{p,i} \pdiff{T_i}{t} &= -\sum_{j}  -\frac{A_{ij}\kappa_{ij}(T)}{V_i} (\grad T)_{ij}  + \int_{V_i} \frac{q_i}{V_i} dV\\
 &=  -\sum_{j}  -\frac{A_{ij}\kappa_{ij}}{V_i} \frac{T_j-T_i}{\Delta x_{ij}}  + \int_{V_i} \frac{q_i}{V_i} dV \label{eq:Teq}
\end{align}
which, after time discretization, yields an algebraic system to solve for the new particle temperatures $T^{n+1}_i$.
The thermal conductivity $\kappa_{ij}$ is calculated as the average of the adjacent particle's temperature-dependent $\kappa(T)$.
The heat source $q_i$\footnote{One can also think of the heat source as an imposed flux condition on the free surface of the finite volumes.} is assumed to only act on the particle surface $S$ based on the furnace temperature $T_f$ and thus is written as
\begin{align}
 \int_{V_i} \frac{q_i}{V_i} dV &= \frac{S_i q_i}{V_i}\\
 q_i &= q_c + q_r\\
 q_c &= h(T_f - T_i)\\
 q_r &= \epsilon \sigma (T_f^4 - T_i^4)
\end{align}
which takes into account heating by convection and radiation, with the convective coefficient $h$, the emissivity $\epsilon$ and the Stefan-Boltzmann constant $\sigma$.
It is assumed that particles ``inside'' of the green body do not experience significant heating by these contributions, similar to \cite{Teixeira2021}.
Instead of determining the set of ``inside'' particles a priori however, the particle's effective surface area $S_i$ is determined during the simulation by a method which assigns zero weight to ``inside'' particles and non-zero otherwise.
This is done by shooting rays from the boundary of the domain until they hit a particle $i$, at which point $S_i$ is increased.
This method naturally accounts for corner and edge effects, as particles located there will have a larger surface area $S_i$ compared to those located on surfaces.
The effects of grain growth and grain motion on the heating are also naturally captured with this method.

Given the number of assumptions in the above, a sanity check for the applicability is warranted.
For this the Biot number $\Bi{}$ is employed, which relates the contributions due to internal fluxes to the heating at the surface of a body.
If  $\Bi{}<0.1$, then the body can be well-approximated as isothermal.
This generally holds for the materials and conditions encountered in sintering as shown in the supplementary material.

As particles used in sintering are commonly on the nanometer to low micrometer scale whereas green bodies are in the millimeter and up range, there still exists a scale gap.
This is especially relevant for temperature, as by the extensive nature of heat, green bodies of different sizes will heat up differently in the same furnace.
Similar to \cite{Teixeira2021}, we bridge this scale gap by introducing a separate length scale factor $\alpha_L$ for the temperature problem only.
For the temperature evolution, all length-related quantities (distances, areas, volumes) are scaled by $\alpha_L$ or an appropriate power thereof.
This effectively scales up the particles and the green body for the temperature evolution, but leaves the microstructure at the smaller scale.
One can think of this as an in-situ multiscale model, in which the microstructure at the original scale gives the geometric details for the temperature evolution at the part scale.

Introducing the scale also allows for easy probing of different Biot numbers in order to verify that disregarding internal temperature gradients is justified.
For this, an even simpler temperature model is employed, which models the green body as a single large body of volume $V$ and surface area $S$, which are the sums over the particle-wise values.
This lets us write the temperature evolution as
\begin{align}
 c_p \pdiff{T}{t} &= \frac{Sq}{V} \label{eq:Teq_hom}
\end{align}
with the heating terms and the heat capacity as before.
If the temporal evolution between the particle-scale and the green-body scale model lie closely together, then the green body's internal temperature gradients are negligible.
Transitively, if the entire green body's temperature gradients are negligible, so are the gradients on a particle scale.
Given the ratio of particle size to green body size for the present study, the Biot number for the green body is about 30 times that of the particle and thus we can estimate when the particle scale temperature model would under-resolve the temperature field.
Also, in order to probe the effects of given heating profiles when the green body can be assumed to be isothermal, the temperature can also be specified as a function of time $T(t)$.

Note that while the model's time evolution is primarily based on convective and radiative heating terms, the heating rate can be adjusted to match e.g. an external surface temperature measurement curve.
This would only involve adding a state-dependent heating term to the surface particles.
Thus the model is also applicable to other high heating rate sintering methods as long as the heating is restricted to the surface of the green body.
Methods which also heat the interior, e.g. by microwave heating, can in principle also be modeled on a particle basis, as long as the resulting heat input does not impose significant temperature gradients on the sub-particle scale.
If it is not, the assumption of isothermal particles could not be held.
But given that these methods should naturally penetrate the body, their spatial variance on the particle scale is likely to be small.
The heating terms could then be formulated as volume integrals over a power density, given either analytically or via a parallel simulation of the target process.

\subsection{Computational notes}
The grid-based PDEs \cref{eq:pfeq,eq:ceq} are solved on a uniform Cartesian grid with finite difference methods.
For the non-advective terms, a simple forward-time-centered-space scheme is applied, with divergences being resolved on cell faces as to ensure conservation.
The advective terms are discretized with a WENO-Z scheme \cite{Borges2008} of order 5 in space and then integrated with a low-storage Runge-Kutta scheme of order 3\cite{Williamson1980}.
By virtue of the advective fluxes being formulated on the cell faces, conservation is also ensured for any advective contribution.
The updates are computed on general-purpose graphics processing units (GPU) employing compute kernels.
The compute kernels are divided at compile-time into boundary-near and inner kernels, with the former running in advance followed by boundary condition application as to improve parallel scaling.
The integrals used in both motion and temperature modeling are largely calculated on the GPUs.
The code is parallelized with the Message Passing Interface (MPI) and thus can use any number of GPUs.

When using more than one GPU, the integrals need to be reduced prior to their use:
For phase-wise integrals, this is a simple global reduction of size $n$.
For pair-wise integrals, this would require too much memory and communication and thus only existing pairs are integrated and then reduced via one-sided communication.
Afterwards, PETSc\cite{petsc-efficient,petsc-user-ref} is employed to build and solve \cref{eq:velsystem} in a least-squares sense and \cref{eq:Teq} as a nonlinear time integration problem using the first-order backward Euler method.
In contrast to \cite{Seiz2023b}, in which a global functional shape was assumed to solve \cref{eq:velsystem} with little computational effort, in the present work the full system is solved.
This is mainly motivated by the fact that once temperature is no longer homogeneous, a simple linear function cannot represent the velocity profile anymore.
Furthermore, since the evolution equations \cref{eq:pfeq,eq:ceq} are now solved on GPUs, the size of the contact matrix $\contact{}$ per worker is increased sufficiently that communication is no longer as much of a bottleneck.

Due to the comparatively small time step and thus small change in state, it is possible to use the integrals of the prior time step as input for \cref{eq:velsystem,eq:Teq}.
This allows submitting compute kernels to the GPU and afterwards computing the solution to \cref{eq:velsystem,eq:Teq} on the central processing unit (CPU), utilizing both the GPU and CPU in parallel.
Ideally, it would be possible to step through single iterations during the solution process and check if another kernel can be submitted or boundary conditions need to be applied, i.e. with the multi-stage Runge-Kutta scheme employed for advection.
At present however PETSc does not offer such a feature.
Thus the work on the CPU can only be hidden completely if it takes less time than the execution of one batch submission of work on the GPU.

The determination of the effective surface area for the particle-wise heating model is similar to a ray tracing problem, though stopping after the first intersection.
As the geometry is only available implicitly from the voxelized phase-field, this is simplified as follows:
For each cell, the phase-field index $i$ of the phase-field with the largest phase-field value is determined.
This yields a scalar field ranging from 0 to $n-1$; 0 represents the vapour phase here.
If a ray traced from the boundary encounters a non-zero scalar, it stops.
The scalar field value at that position serves as the index of the particle surface area $S_i$ to increment.
Furthermore, the neighboring values perpendicular to the ray direction are considered to improve the area estimate:
If these are of the same index, they are also counted towards the surface area.
This would lead to overcounting by a factor of $f_d = 2(d-1)+1$, with the dimension $d$ of the simulation domain, on average and thus the resulting sum is divided by this factor.
For simple shapes shapes (cubes, spheres) after coarse voxelization this yields the surface area to within about $10\%$ error.
By splitting the particle surface area into components per view-direction, it is also possible to introduce directional heating.

Due to the strongly variable memory accesses required for this, the algorithm is not necessarily suited for GPUs.
Rather, it is implemented on the CPU and employs a dynamic bounding box to reduce the necessary number of memory accesses.
Since the scalar field these calculations are based on only changes quite slowly, this algorithm is executed in the background at least every $k$ steps, including asynchronous MPI communication by the main thread.
Early testing showed that the algorithm is finished after less than 10 steps for typical geometries; though since practically the scalar field changes much more slowly, $k=500$ is chosen.
As it is done entirely asynchronously using mainly CPU resources, the algorithm has close to zero impact on the runtime of the simulation.

\subsection{Measurement methods}
The density $\rho$ is computed by a geometric approach:
First, the bounding box of the green body is found via the extremal positions of cells for which $\phi_v < 0.5$.
This bounding box generates smaller bounding boxes by moving the edges of the bounding box in increments of $\frac{1}{8}$ of the overall bounding box, resulting in a series of bounding boxes.
For each bounding box the integral $\rho = \frac{1}{V_b}\int \phi_s dV$ is computed with the respective bounding box volume $V_b$ and the solid phase-field $\phi_s$.
As the outermost bounding box includes strong surface effects, it is not used, but rather the first inner bounding box representing about $(\frac{7}{8})^3$ of the green body.
By virtue of the initial bounding box determination, if surface diffusion causes the outer particles to temporarily elongate (as e.g. shown by \cite{Djohari2009a} and others), the measured density may temporarily decrease.

The average grain size is based on the particle volumes $V_i$ which are simple volume integrals.
The equivalent grain-size based on a sphere is calculated and the volume-weighted average of these is taken to represent the average grain size.

%% file: parts/results.tex
For the initial conditions the particle packing also used in \cite{Seiz2024} is employed.
This packing consists of 3431 particles of radius \SI{12}{nm}, forming a cubic green body with an edge length of about \SI{380}{nm}.
No-flux conditions are applied on all global boundaries.
The particles are initialized via planar equilibrium phase-field profiles in the shapes of spheres, followed by an equilibration to allow for adjustment to curved interfaces and for the initial formation of grain boundaries.
For this process the GB mobility is artificially reduced to be insignificant, with both interfacial diffusivities set to zero and $D_b = \SI{1e-12}{m^2/s}$

\begin{figure}[h]
\centering
 \includegraphics[width=0.5\textwidth]{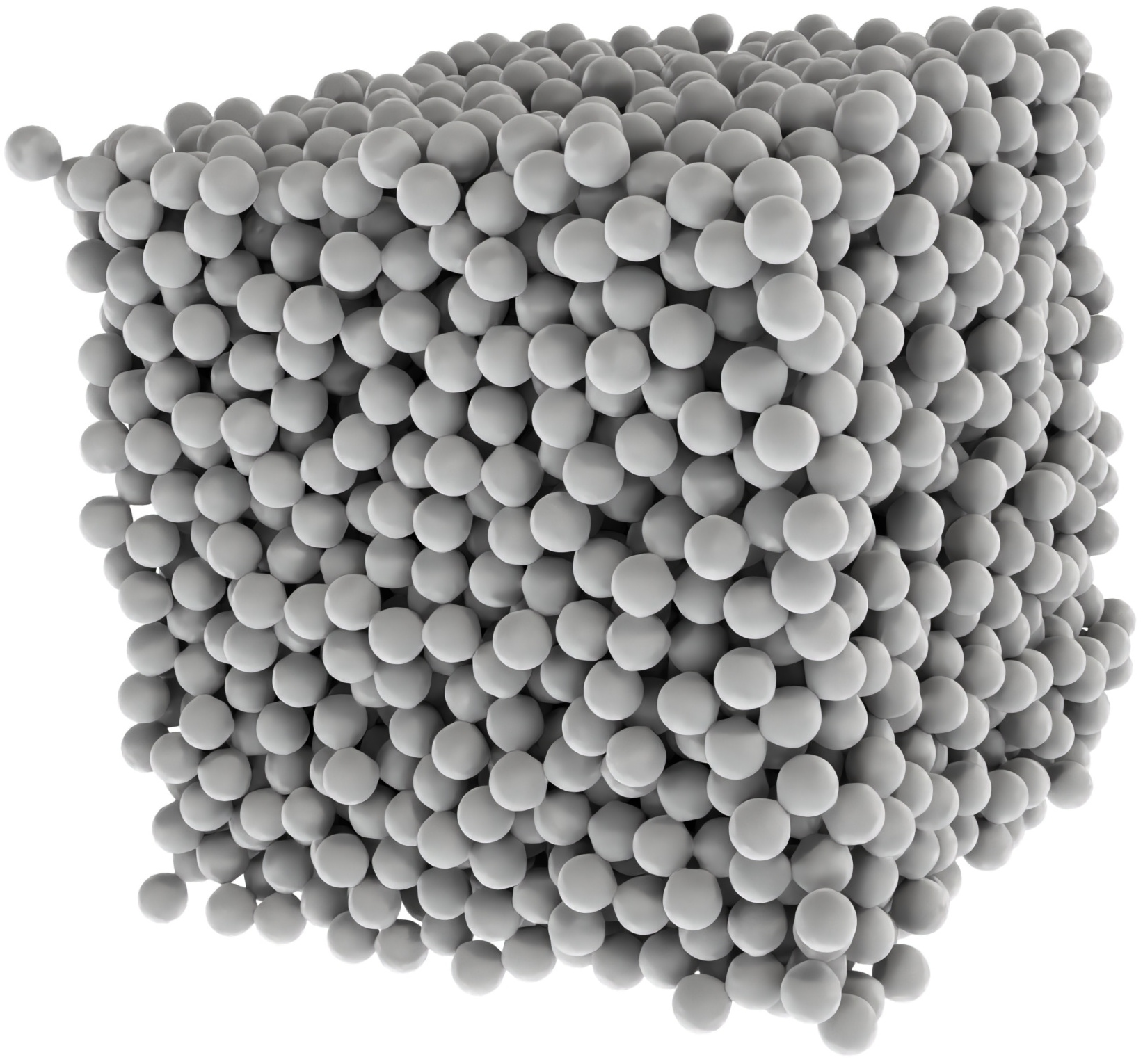}
 \caption{The initial green body configuration is depicted.}
 \label{fig:initial}
\end{figure}

The green body's initial state is depicted in \cref{fig:initial}.
For all visualizations of the microstructure, ParaView\cite{ParaView} is used.
When presenting the green body without a color map, the maximum non-vapour phase-field $\phi_m$ is used to produce a contour surface at $\phi_m = 0.6$, which eases differentiation of grains and obtains an etching-like effect at triple and higher order grain junctions.
For all plots, matplotlib's\cite{Hunter:2007} Python interface is used.

The first investigation is concerned with whether effects associated with rapid-heating can be reproduced by employing the simplest temperature model of a given function $T(t)$.
There are two qualitative observations we would like to confirm:
First, the density observed at a fixed temperature decreases as the heating rate is increased\cite{Wang2020}.
This is simply due to less time being available for diffusive processes to act.
Second, if the activation energy for densification is higher than for grain growth, then at the same density, a lower grain size should be obtained with higher heating rates.
This is, again, due to less time being available for grain growth to reduce the driving force for densification before densification itself starts using the driving force.
Assuming that the activation energy for grain growth is given via the GB mobility (\SI{443}{kJ/mol}) and that for densification is given via the GB diffusion coefficient (\SI{825}{kJ/mol}), then the present system should show such behaviour.

The second investigation is concerned with inhomogeneous heating effects.
Here it is explored when inhomogeneous effects on the microstructure become evident and whether this can be predicted by the Biot number.
This includes a verification of the particle-scale model by comparing its predictions with the homogeneous temperature model.

In all cases, the initial particle temperature is $T_i=\SI{300}{K}$, with the end temperature being $T_e=\SI{1800}{K}$, taken on the lower end for sintering temperatures of alumina.
If the temperature is specified as a function, this implies a linear ramp up to $T_e$ followed by holding the temperature.
Otherwise, the furnace temperature $T_f$ is set to $T_e$, which implies an evolution towards $T_e$ due to the structure of the heating terms.
For simplicity, the end temperature is generally held; realistic heating schedules would include a cooling ramp.
Simulations are continued until at least $95\%$ density is reached.
Videos of the simulations depicting the entire evolution are deposited with the supplementary material.

\subsection{Controlled heating}
The first study employs constant heating rates $\dot{T} \in \{50, 125, 250, 500\}\si{K/s}$ as a broad sweep across heating rates.
Additionally, instantly reaching $T=T_e$ is also considered as to determine whether \SI{500}{K/s} is practically equivalent to instant heating.
The temperature in the domain is thus a simple function of time $T(t)$.
The range of heating rates is chosen to be in the range of heating rates achievable by modern high-heating rate experiments\cite{Wang2020,Porz2022a}.
Heating rates as in conventional sintering on the order of \SI{1}{K/min} are not considered, as their simulations would require an excessive amount of time steps with a forward Euler method.
An estimate for the number of time steps dependent on the heating rate is given in the supplementary material; as a simple example, it would take about \num{33e7} steps to reach $T_e$ with a heating rate $\SI{0.5}{K/s}$.
This number of steps would correspond to about half a year of computation, including queueing times, which is deemed excessive.
Future work will consider alternative time integration schemes such that lower heating rates become accessible with acceptable computational effort.

\begin{figure}[h]
\centering
\includegraphics[width=\textwidth]{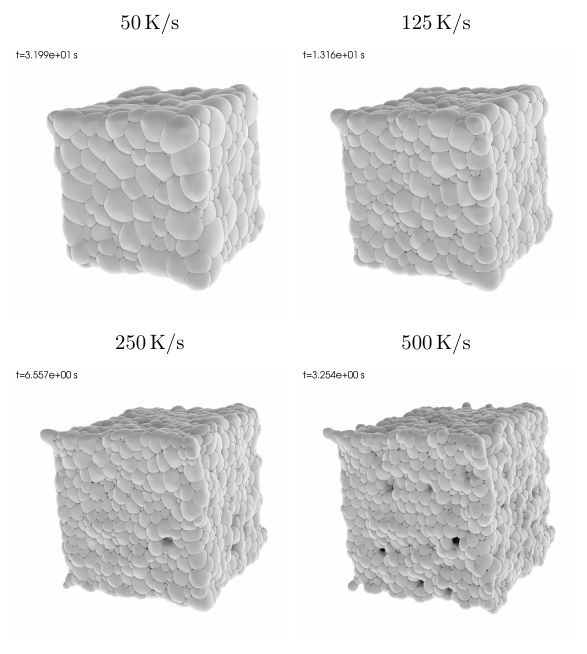}
 \caption{Exemplary microstructures at approximately $90\%$ density are depicted. Note the much larger grains for slower heating rates.}
 \label{fig:hom-microstructure}
\end{figure}

Examples of the resulting microstructures at approximately $90\%$ density are shown in \cref{fig:hom-microstructure}.
Here one can easily identify much larger grains for simulations with slower heating rates.
There also seems to be a trend for the surface to still possess open porosity with faster heating rates; this would be beneficial in atmospheric sintering as to allow gas to escape.
These regions visually tend to have the highest number of surrounding grains and thus their elimination is facilitated by grain growth, as it reduces the coordination number.

\begin{figure}[h]
 \centering
  \begin{subfigure}[t]{0.75\textwidth}
    \includegraphics[width=\textwidth]{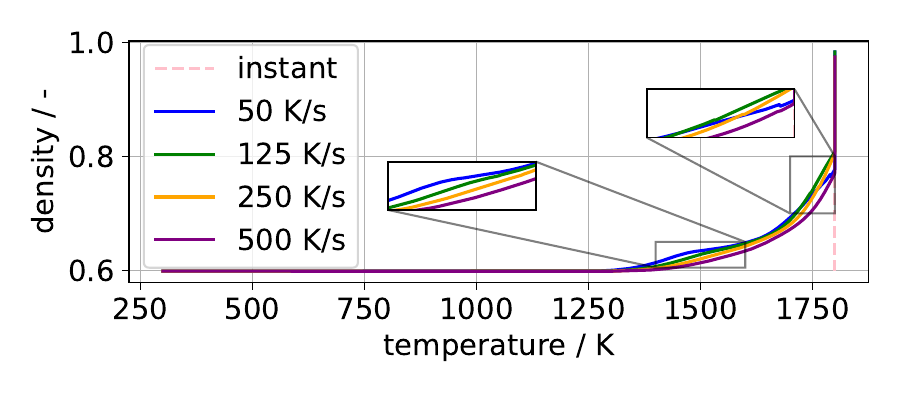}
    \caption{temperature-density trajectory}
    \label{fig:constrate-Trho}
  \end{subfigure}
  \begin{subfigure}[t]{0.75\textwidth}
    \includegraphics[width=\textwidth]{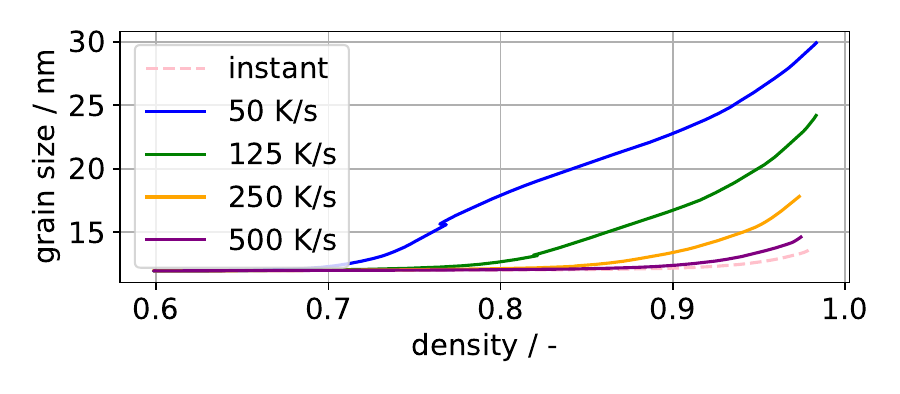}
    \caption{sintering trajectory}
    \label{fig:constrate-grho}
  \end{subfigure}
  \caption{Two trajectory types for high-heating rate sintering at a constant heating rate and homogeneous temperature are shown. The qualitative trends of experiments are generally captured, except for the lowest heating rate showing a crossover with higher heating rates at higher density.}
\end{figure}

The temperature-density trajectory of the simulations is depicted in \cref{fig:constrate-Trho}.
Initially, the trend for higher heating rates exhibiting a lower density at equivalent temperature is kept, cf. the leftmost inset.
The lowest heating rate exhibits a kind of plateau-effect at low densities ($\approx 63\%$), after which it follows a convex trajectory like the other curves.
This causes its trajectory to intersect with the two middle heating rates before reaching $T_e=\SI{1800}{K}$ (cf. rightmost inset), contrary to expectations.
The remaining trajectories however all show the experimentally observed ordering, i.e. the higher the heating rate, the lower the trajectory is on this plot.
While keeping this in mind, consider the sintering trajectories depicted in \cref{fig:constrate-grho}:
Here the expectations for high-heating rate sintering are fulfilled, with higher heating rates allowing for smaller grain size at the same target density.
It can also be observed that with the present materials parameters, even a heating rate of \SI{500}{K/s} is not exactly comparable to instant heating, even though it would be considered practically instant in experiments.
Furthermore, grain growth starts before the final sintering stage is entered ($\approx 90\%$ density).
This is due to two reasons:
First, nanocrystalline materials tend to show grain growth prior to the final sintering stage\cite{Chen2000b}, both due higher driving forces for grain growth (modeled) and potentially different grain growth mechanisms \cite{Krill2001} (not modeled).
Second, by virtue of the finite extent of the domain, particles on the outside will experience a different capillary pressure compared to those on the outside.
Thus, even though the simulation is started from an almost perfectly uniform size distribution, there is a systematic disturbance which leads to earlier grain growth.
Note that the lowest heating rate does not show significant grain growth during its plateau regime; thus grain growth cannot explain the plateau.

\begin{figure}[h]
 \centering
  \begin{subfigure}[t]{0.7\textwidth}
    \includegraphics[width=\textwidth]{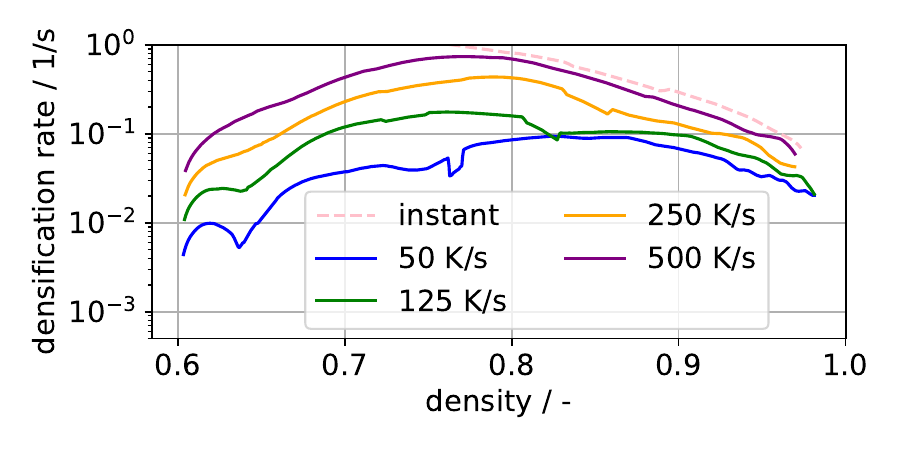}
    \caption{simulation results (this work)}
    \label{fig:dens-densirate}
  \end{subfigure}
  \begin{subfigure}[t]{0.7\textwidth}
    \includegraphics[width=\textwidth]{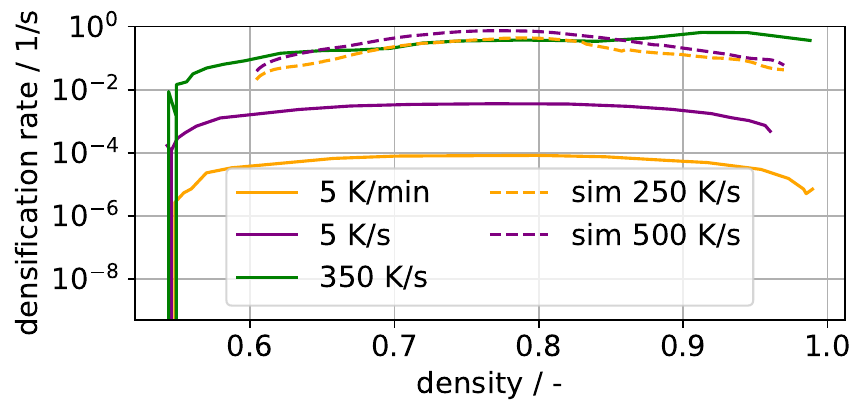}
    \caption{experimental results (\cite{Wang2020})}
    \label{fig:dens-densirate-expLit}
  \end{subfigure}
  \caption{Density-densification rate trajectories for the present simulations as well as experimental data from literature. The trajectory is generally concave, showing its maximum around $80\%$ density. The experimental curve for \SI{350}{K/s} is of the same order-of-magnitude as the simulations which bound this heating rate.}
\end{figure}

Another compact way to show the effects of high-heating rate sintering is a plot of densification rate over density.
This is shown for the present study in \cref{fig:dens-densirate} and below it for the experimental results of \cite{Wang2020}.
\cor{
Within these experiments, the temperature range of about \SIrange{300}{2000}{\kelvin} was explored with different heating techniques and resulting heating rates.
}
The experiments tend to show a concave curve, with the maximum densification rate occuring around $80\%$ density, with higher heating rates pushing the curve upwards.
For the heating rate of $\SI{350}{K/s}$ it is likely that the temporal resolution of measurements was insufficient to resolve the curve around the maximum densification rate and beyond and thus it shows rather several broad maximum regions.
Broad qualitative features are retained in the present simulations, i.e. the concave shape, the maximum at around $80\%$ density and the effect of heating rate.
For the experimental heating rate $\SI{350}{K/s}$ that is bounded by the simulation heating rates, the densification-density trajectory is bounded by the simulative results for a significant span of data, without any attempt being done to match coefficients, as shown by the dashed curves.
This suggests that the present model is sufficient to estimate heating schedules for target densities.
The earlier plateau is transformed into an early decrease of densification rate here.

\begin{figure}[h]
\centering
  \begin{subfigure}[t]{0.4\textwidth}
    \includegraphics[width=\textwidth]{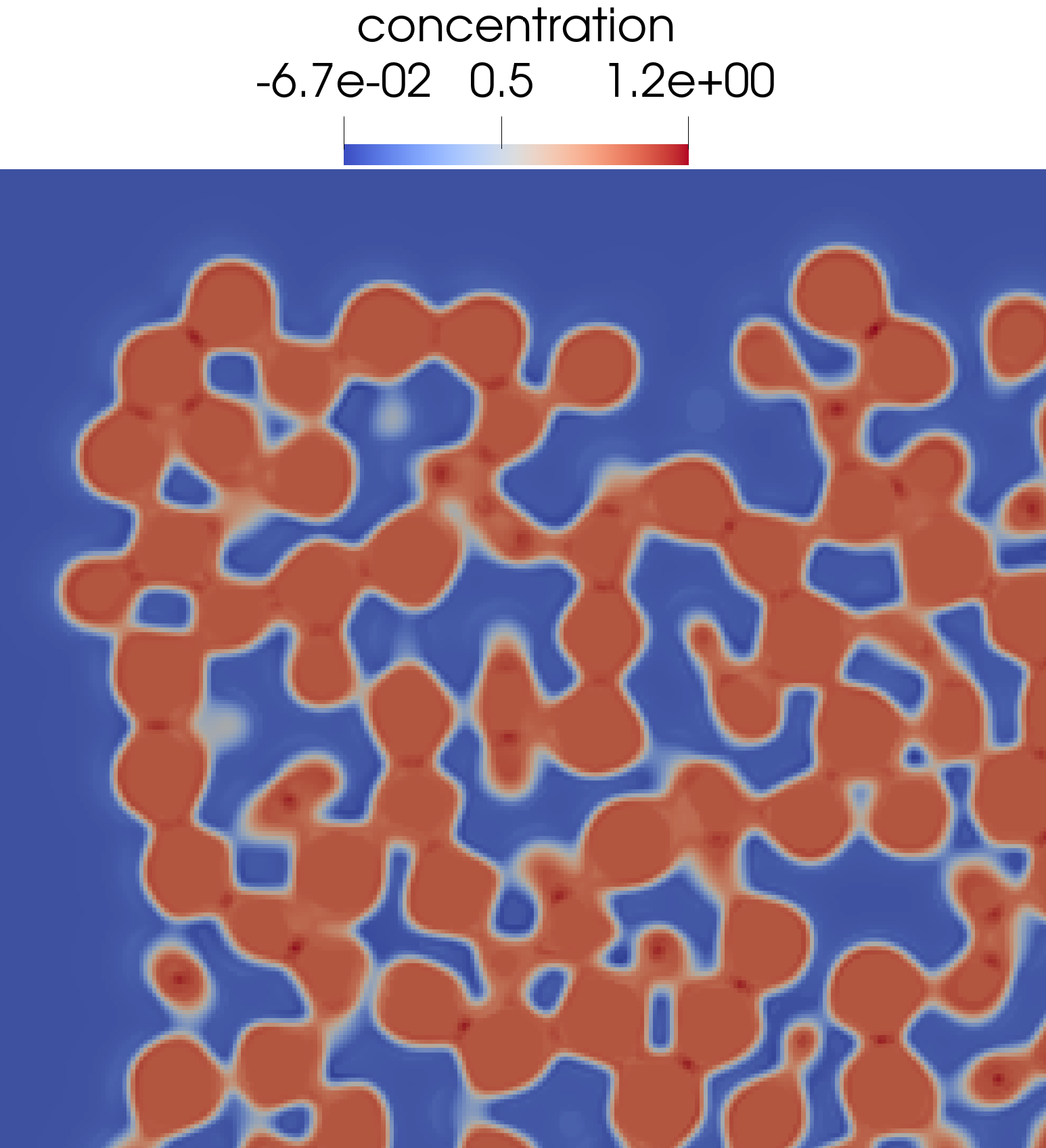}
    \caption{$\dot{T}=\SI{50}{K/s}$}
  \end{subfigure}
  \begin{subfigure}[t]{0.4\textwidth}
    \includegraphics[width=\textwidth]{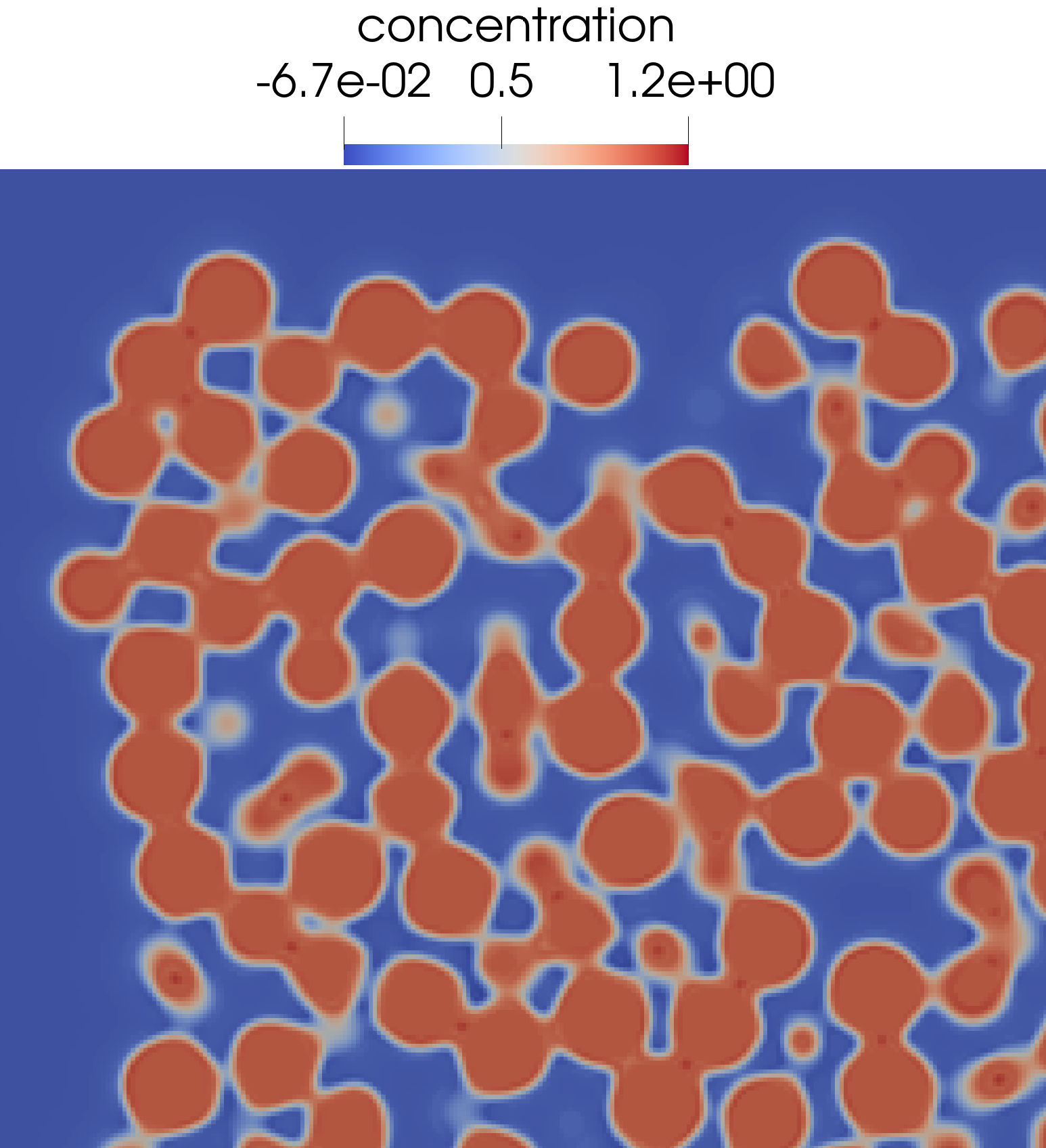}
    \caption{$\dot{T}=\SI{500}{K/s}$}
  \end{subfigure}
 \caption{Solid concentration field at about $63\%$ density for two heating rates. The lower heating rate shows larger and longer-lasting accumulation of solid concentration on the grain boundary, causing a plateau in the densification curve.}
 \label{fig:microstructure}
\end{figure}

As the averaged properties of the microstructure did not deliver an explanation for the plateau behaviour, the microstructure itself is compared next.
A close-up of a planar cut of the solid concentration fields\footnote{Note that the concentration is not bounded by $[0,1]$ as the free energy \cref{eq:free-energy} is a parabolic approximation.} is shown in \cref{fig:microstructure} for heating rates $\in \{50, 500\}\si{K/s}$ at approximately the same density of $63\%$.
The color scale is based on the limits of the simulation with $\dot{T}=\SI{50}{K/s}$ at this point in time.
It can easily be seen that the lower heating rate has a much larger accumulation of solid on the grain boundaries than the higher heating rate.
This accumulation is caused in the first place by the compressive velocity profile across grain boundaries, which implies some amount of accumulation.
Since the absorption dynamics (\cref{eq:vacabsorb}) depend on the difference of the local concentration and an approximated equilibrium concentration $c_{eq}(\mu_\alpha\beta)$ of the grains forming the GB, such accumulation will significantly affect the dynamics.
Given that $c_{eq}(\mu_\alpha\beta) = c_{0,s} + \frac{\mu_s}{k}$ is based on the surface chemical potential $\mu_s \approx \gamma/r$ and with the value of $k=\SI{5e9}{J/m^3}$ employed in this work, $c_{eq}(\mu_\alpha\beta)-c_0 \leq 0.1$, i.e. significantly below the observed concentration accumulation of up to 1.2 $(c-c_0 = 0.3)$.
Thus this accumulation leads to a temporary slowdown of densification dynamics, as GBs with this accumulation can contribute de-densifying terms to the system \cref{eq:velsystem}; note however that since not all GBs show such an accumulation that the net solution is still densifying.
At higher heating rates GB diffusion is already sufficiently activated as to redistribute the accumulation on a similar timescale as it occurs, at equivalent density.

Since this effect was only observed for a single simulation, with the simulative trends following experimental trends otherwise, the present paper will only list a few possible remedies:
The most obvious is one is projecting the velocity field to a divergence-free space, as commonly done in incompressible fluid dynamics.
This would involve calculating a divergence and a gradient as well as solving an additional Poisson equation at each step, with the latter likely taking longer than the rest of the solver combined if forward Euler time stepping is employed.
Another method could be, given that the gradient of chemical potential is sufficiently smooth in the grain boundary, to direct the advective flux along the gradient of chemical potential in the grain boundary.
This method would resolve the accumulation by transporting it quickly along the grain boundary to the sintering neck, similar to the action of GB diffusion at higher heating rates.
It would only involve gradient calculations and local interpolations and thus be comparatively cheap.
A separate work investigating both of these methods is planned.

Sintering models based on solving the incompressible Navier-Stokes equations accounting for microstructural effects such as \cite{Yang2025} would typically not exhibit the accumulation effect given the assumption of incompressibility.
However, these models have, to the authors' knowledge, only been used for the computation of viscous sintering, which does not involve GBs as necessary sinks for vacancies.
Whether these models can be extended to account for vacancy sink activity is another point for future work.
Using such kinds of models also requires the use of time integration schemes with less restrictive stability bounds as to compensate for the high computational cost of solving the incompressible Navier-Stokes equations.

\subsection{Uncontrolled heating}
In this study, the green body ($T=T_i$) is inserted into a pre-heated furnace at $T_f=T_e$, which causes the green body to heat up by convection and radiation.
Two models for the temperature are employed and compared herein, the particle-resolving, inhomogeneous model \cref{eq:Teq} and the homogeneous model \cref{eq:Teq_hom}.
Both models are calculated for three different thermal length scale factors $\alpha_L \in \{\num{1e3},\num{5e3},\num{1e4}\}$, corresponding to initial scaled green body sizes of $L \in \{0.38, 1.9, 3.8\}\si{mm}$ respectively.
The total Biot number of the green body, estimated at $T=T_f$, is about $\num{0.017}$, $\num{0.085}$ and $\num{0.17}$ respectively.
This indicates that one should not see significant differences between the models for $L=\SI{0.38}{mm}$, but they should start appearing for the larger two green bodies.

\begin{figure}[h]
\centering
\includegraphics[width=0.45\textwidth]{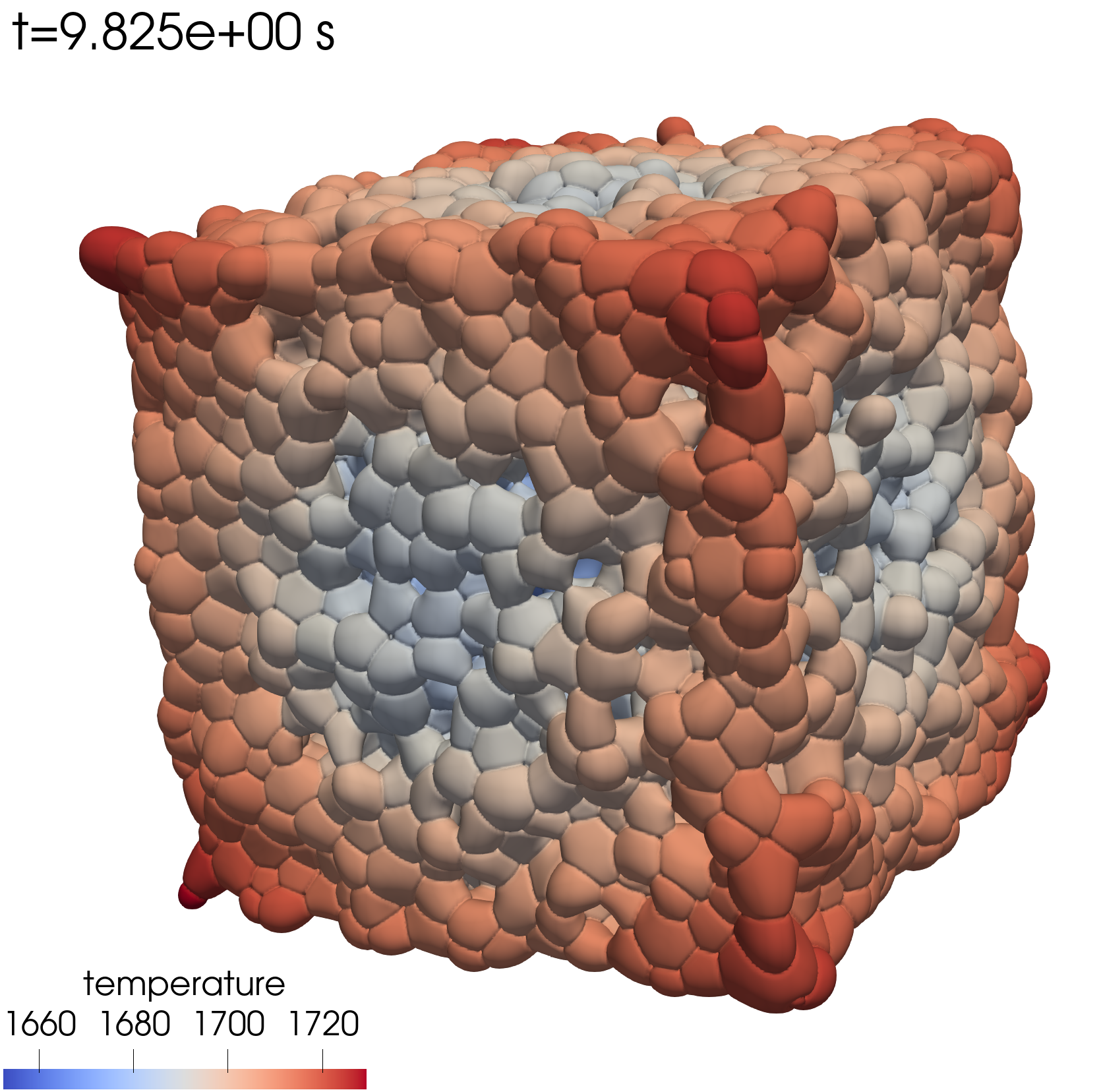}
\caption{Temperature distribution at $70\%$ density for $L=\SI{3.8}{mm}$ and a particle-wise resolved temperature field. By virtue of the edge effect, corners and edges heat up faster than faces and the interior of the green body is cooler still.}
\label{fig:dem-temperature-distri}
\end{figure}

\begin{figure}[]
\centering
\includegraphics[width=\textwidth]{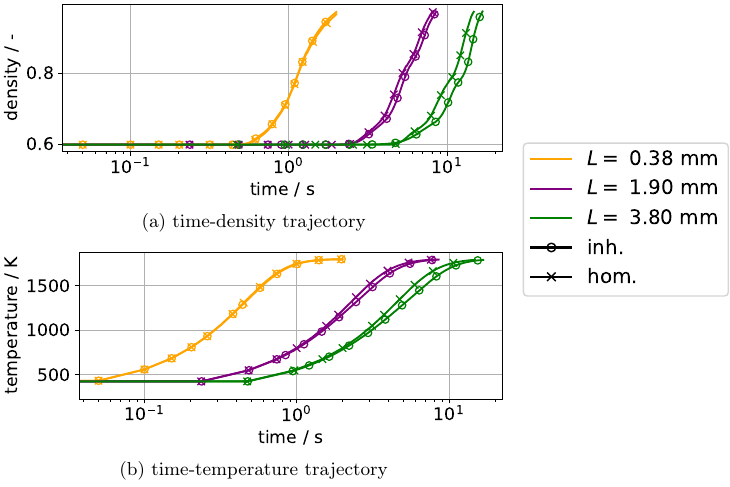}
 \caption{Time and temperature trajectories for uncontrolled heating are depicted. The size of the green body $L$, and thus the Biot number, gives a good indication of when the inhomogeneity of the temperature starts to affect the evolution, i.e. when the markers of the same color do not match each other.}
 \label{fig:uncontr-t}
\end{figure}

For inhomogeneous heating, a certain spatial profile develops.
As an example of this, the temperature distribution is depicted in \cref{fig:dem-temperature-distri} at about $70\%$ density for $L=\SI{3.8}{mm}$:
The corners and edges of the cubic green body exhibit the highest temperature, with the middle of the faces being cooler, and the inside of the green body even cooler.
This is to be expected simply due to edge effects which are naturally captured by the heating model.
The temperature distribution is qualitatively similar independent of the green body size, with the main effect of green body size being the temperature difference scale.
At this stage the temperature difference between inside and outside is only about \SI{80}{K}.
In the early stages this temperature difference can be up to \SI{1000}{K}, which would imply significant thermal stresses if they were modeled.

In \cref{fig:uncontr-t}(a) the density-time trajectories of the simulations are plotted.
Here one can observe that indeed $L=\SI{0.38}{mm}$ shows little difference between either model and that once the Biot number is large enough, the microstructural evolution of the green body is affected, with the density-time curves shifting ever wider apart with increasing Biot number.
A similar trend is observed in the temperature-time curve depicted in \cref{fig:uncontr-t}(b), with the average temperature differing substantially for higher Biot numbers before closing in on $T_f$.
This also shows the extensive effect of heat:
The larger the green body, the longer it will take a furnace at a fixed temperature to heat it to a certain temperature, as the heat input does not scale linearly with the total heat capacity of the green body.
Based on these observations, it is confirmed that the Biot number is a good predictor for temperature inhomogeneity for a complex microstructure in a furnace.
For the present case we can quantify $\Bi{} \ll 0.1$ for approximate temperature uniformity as $\Bi{} \leq \num{0.017}$.

\begin{figure}
\includegraphics[width=0.9\textwidth]{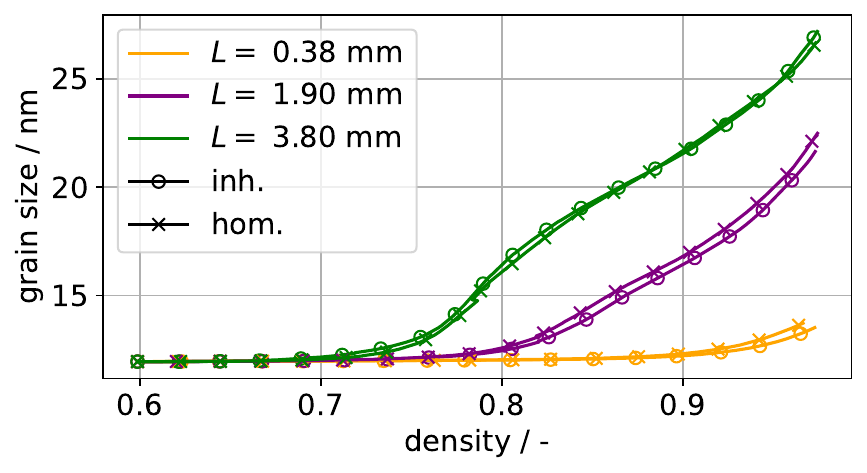}
\caption{Sintering trajectory for uncontrolled heating. Since smaller green bodies sinter more quickly, they also exhibit lower grain sizes at the same density. The variation in trajectory between the models is not a simple function of the green body size.}
\label{fig:uncontr-trajectory}
\end{figure}

\begin{figure}[]
\centering
\includegraphics[width=\textwidth]{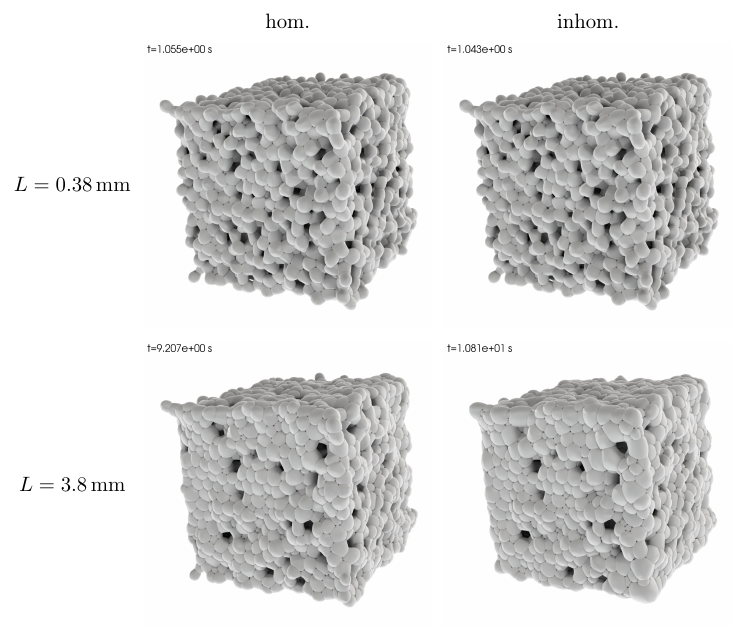}
 \caption{The microstructures at approximately 75\% density are shown for various simulations. There is hardly any difference between the temperature models for the smaller green body $L=\SI{0.38}{mm}$, but a significant one at $L=\SI{3.8}{mm}$: Both the corners and edges have started grain growth.}
 \label{fig:inhom-microstructures}
\end{figure}

The sintering trajectories of these simulations is shown in \cref{fig:uncontr-trajectory}.
As expected, a smaller green body allows for a smaller grain size at the same density due to heating more quickly.
However, all curves show a difference between the two heating models, regardless of the Biot number.
This can be attributed to inhomogeneity in the temperature leading to microstructural inhomogeneity.
Interestingly, the simulations with the largest Biot number and thus inhomogeneity do not show the largest difference in trajectory.
In order to explain this, a look at the microstructure is necessary.
These are exemplarily shown in \cref{fig:inhom-microstructures} at approximately 75\% density.
First, note that at the smallest Biot number ($L=\SI{0.38}{mm}$) there is hardly any difference in microstructure.
Second, when increasing the Biot number, a change in the microstructure is observable:
The surface tends to look more dense via the same mechanism as for high heating rates, i.e. grain growth reducing pore coordination number.
Finally, at higher Biot numbers, the inhomogeneous temperature simulations tend to already show grain growth starting from the corners and edges as these exhibit the highest temperatures. (cf. \cref{fig:dem-temperature-distri}).
Thus the lack of divergence \emph{on average} is likely the simple result of the average hiding the difference between the simulations; one could have giant grains on the outside with the same structure inside and obtain a similar average grain size by virtue of the surface to volume ratio.

\begin{figure}[]
 \centering
  \begin{subfigure}[t]{0.45\textwidth}
    \includegraphics[width=\textwidth]{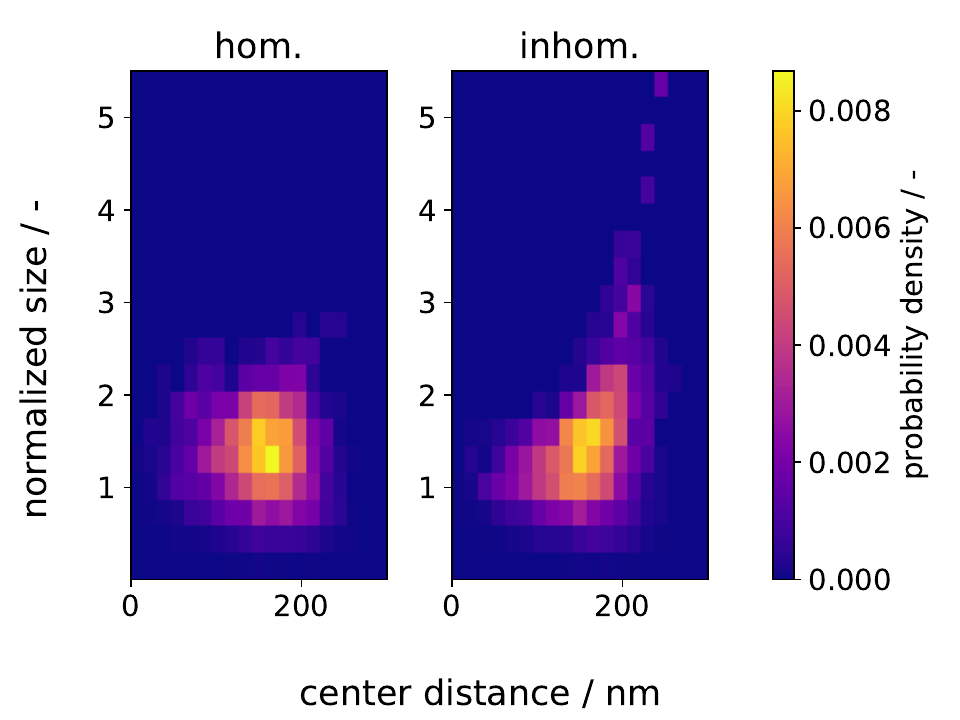}
    \caption{75\% density}
    \label{fig:heatmap-gs-075}
  \end{subfigure}
  \begin{subfigure}[t]{0.45\textwidth}
    \includegraphics[width=\textwidth]{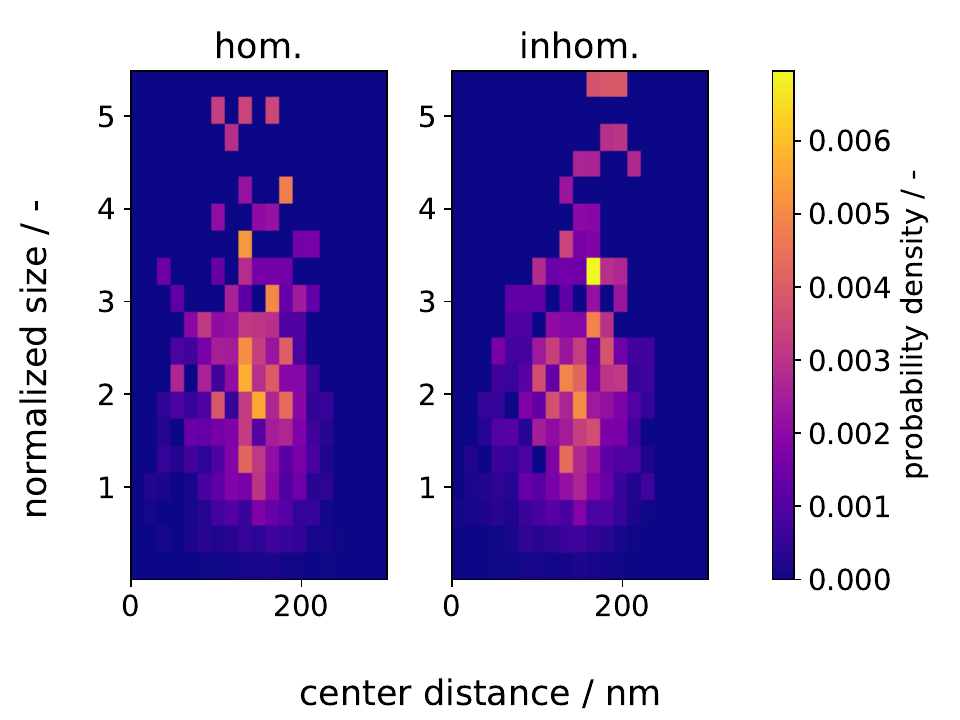}
    \caption{90\% density}
  \end{subfigure}

 \caption{Heatmaps of center distance to normalized grain size for $L=\SI{3.8}{mm}$ for two densities are shown. At lower densities, the microstructural spatial inhomogeneity due to temperature inhomogeneity is visible, but this effect is lost at higher densities due to the sintering front having advanced through the green body.}
 \label{fig:heatmap-gs}
\end{figure}

In order to differentiate microstructures with similar averages but spatial differences, heatmaps of the $i$-th grain center distance $d_i = ||\vec{p}_i - \vec{c}||$ and its volume normalized by the current average are depicted in \cref{fig:heatmap-gs}:
At 75\% density one can see a clear difference in the distribution between the temperature models, with a more or less uniform distribution for homogeneous temperature, but a distribution which is slanted to larger grain sizes at larger center distances for inhomogeneous temperature; this is precisely what is observed in \cref{fig:inhom-microstructures}.
However, at larger densities this difference tends to vanish again.
This is likely due to the formation and motion of a sintering front\cite{Garcia2011}:
Sufficiently hot grains will activate densification and grain growth on the neighboring GBs, with local densification occurring until these encounter either too cold grains or the driving force has been spent.
This process also substantially increases the GB area and thus the heat transfer to colder particles inside, which are then in turn activated to densify and undergo grain growth.
As long as the front activates the next inner particles before having densified completely, the front can advance homogeneously through the body, leaving behind a similar structure as in the homogeneous temperature case.
The deviation in \cref{fig:heatmap-gs-075} is thus simply due to the front not having advanced yet through the green body.

\FloatBarrier

%% file: parts/acks.tex
This paper is based on results obtained from a project, JPNP22005, commissioned by the New Energy and Industrial Technology Development Organization (NEDO).